# Molecular dynamics study of plasticity in *Al-Cu* alloy nanopillar due to compressive loading

Satyajit Mojumder*

[a] Department of Mechanical Engineering, Bangladesh University of Engineering and Technology, Dhaka-1000, Bangladesh.

**Abstract:**

In this paper, compressive loading effects on the plasticity of *Al-Cu* alloy varying the crystal orientation of *Al* and alloying element (*Cu*) percentage are investigated using molecular dynamics approach. The alloying percentage of *Cu* are varied up to 10% in <001>, <110> and <111> crystal loading direction of *Al*. Present results indicate that the alloy nanopillar has highest first yielding strength and strain along <110> and <001> direction, respectively. Further, the dislocation density and dislocation interactions are studied to explain the compressive stress-strain behavior of the alloy nanopillar.

## 1. Introduction

Nanostructures of metals and alloys such as nanowire, nanoribbon, nanopillar etc. are given prodigious importance due to its wide variety of application in MEMS/NEMS[1]. In many applications, nanostructures are subjected to compressive loading and nanopillars are designed for this purpose[2,3]. Compare to traditional bulk counterpart, the nanostructure materials are more suitable to carry the compressive load due to its enhanced mechanical properties. Aluminum and one of its major alloy with Copper have tremendous potential to be used as nanopillar[4].

*Corresponding author: Tel: +880-1737-434034, E-mail address: satyajit@me.buet.ac.bd

Nanoscale material properties are very important to ensure the suitability of their application. Nanopillars are fabricated in the laboratory and they are indented or compressed with different nanotools for extracting the mechanical properties[5,6]. Previously, pure Al nanopillars[7] are experimentally fabricated and tested for their mechanical properties. Molecular dynamics is a widely used method to investigate the mechanical properties of nanomaterials through computer simulation. Both single crystal of *Al* [8] and *Cu* [9,10] were studied previously under tensile loading using molecular dynamics method to predict their mechanical properties.

Crystal orientation during the loading also has a significant impact on the mechanical properties. This happens due to the orientation of the slip plane in FCC metal. Previously, crystal orientation effects are investigated for the Titanium[11], Magnesium[12], Aluminum[13], Iron[14] nanopillar, etc. The alloys in nanoscale can be a suitable replacement of the pure metals due to its superior mechanical properties. Alloying in a different orientation of a crystal can make an influential role for the suitable crystal slip plane and the mechanical properties are affected significantly by solid solution strengthening. Solid solution strengthening is the dominant mechanism which makes alloys stronger than the base material[15]. When an atom of alloying metal is added to the crystalline lattice of the base metal, it forms a solid solution. The added atom of alloying element creates a strong tension or compression field based on its atomic size compare to the host base metal. This localized strain field interacts with the dislocation and affects the dislocation movement path. Leyson et al. [16] studied the solute-dislocation interaction for *Al* alloy with *Mg, Si, Cu* and *Cr* alloying element and quantitatively assessed the energy barrier produced by the different solute atom. Ma et al. [17] investigated lattice misfit due to the *Cu* solute atom in *Al* base metal and concluded that strengthening capability is highly dependent upon the solubility. Plasticity occurs in the nanopillar when it is compressed afterward of its yielding. The plastic deformation is governed by the dislocation nucleation, propagation, and interaction[18]. The alloying element and its percentage have a strong impact[19] on the dislocation activity and the plastic behavior of the alloy become a more complex phenomenon.

The alloying element percentage and crystal orientation along the loading direction are two of the many important factors, which play a significant role in plasticity of nanopillar. Therefore, the aim of this paper is to investigate these effects on *Al-Cu* alloying nanopillar under the application of compressive loading which is very common for nanostructure.

## 2. Methodology

The uniaxial compression simulations are performed for *Al-Cu* alloy nanopillar having a dimension of 6nm×6nm×12 nm. The EAM alloy[20] potential is used to describe the interaction between *Al* and *Cu*. The aspect ratio of the nanopillar height to width is kept constant as 2:1 and the compression is applied in crystal direction of <001>, <110> and <111> of *Al*. For the alloy modeling, first, the pure *Al* in different orientations are prepared and then *Al* atoms are randomly replaced with *Cu* atom for different weight percentage of *Cu* in *Al*. All the simulations are carried out using LAMMPS[21] software package and OVITO[22] is used for the post-processing purpose. The initial geometry of the alloy is relaxed sufficiently (for 100 ps) under the NPT dynamics. Later, a compressive load is applied varying the *Cu* percentage and crystal orientation of *Al* at a temperature 300K along the negative Z direction (see Fig. 1) of the simulation box for strain rate of $10^9$ s$^{-1}$. The timestep chosen for all the simulations is 1fs. For obtaining the stress-strain behavior, atomic stresses are calculated as the simulation box is deformed uniaxially. Atomic stresses are calculated based on the definition of virial stress [23] as follows

$$\sigma_{virial}(r) = \frac{1}{\Omega}\sum_i \left[\left(-m_i \dot{u}_i \otimes \dot{u}_i + \frac{1}{2}\sum_{j \neq i} r_{ij} \otimes f_{ij}\right)\right] \quad (1)$$

where the summation is over all the atoms occupying the total volume, $m_i$ is the mass of atom $i$, $\dot{u}_i$ is the time derivative which indicates the displacement of atom with respect to a reference position, $r_{ij}$ is the position vector of atom, $\otimes$ is the cross product, and $f_{ij}$ is the interatomic force applied on atom $i$ by atom $j$.

## 3. Results and discussion

The stress-strain diagrams obtained from the compressive loading simulations are presented in Fig 2. The different alloying percentage and the loading directions have a significant impact on the stress-strain curve. The stress-strain curves have two distinct regions in the diagram as the previous study of *Cu* [10] and *Fe* [14] nanopillar.

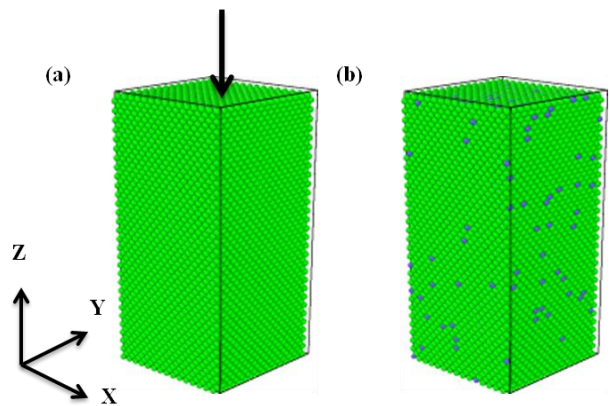

Fig. 1: (a) Crystalline *Al* nanopillar, (b) *Al-Cu* alloy nanopillar with 10% *Cu*. The *Al* and *Cu* atoms are represented by green and blue color, respectively.

The stress increases linearly up to a certain point (the first yielding point) and then start to fluctuate as flow stress. The flow stress for the tensile simulation tends to show very little variation but in compressive loading, the fluctuation can be significant due to the activation of several cross slip system in the material. Previous studies on single crystalline *Al*[8] and *Cu*[10] show the similar type of stress-strain curve what we obtained here, however, molecular dynamics study are not available for *Al-Cu* alloy for a direct comparison purpose. One of the major differences in the single crystal and in the alloy is fluctuation in flow stress. As the dislocation intermittently changes its slip plane, interact with each other, it is more likely that there is more fluctuation in the flow stress region of the stress-strain curve of *Al-Cu* alloy compare to the single crystal *Al* and *Cu*. The stress-strain curve can further be explained by solid solution strengthening mechanism. The lattice constant for *Al* and *Cu* is 4.04 $A^o$ and 3.61 $A^o$, respectively. When an *Al* atom is replaced by *Cu* atom in a certain lattice the lattice misfit creates a compression stress field inside the lattice. Lubarda[24] showed that Cu solute atom in Al causes a size factor reduction of 0.378. Therefore, when the load is applied on the lattice, and dislocation tries to move along the slip plane it faces a strong energy barrier in its forward path. As a result, dislocations choose its energetically favorable slip plane in such condition and require higher strength to move.

In <001> direction compression, the yielding does not show sharp peak as <110> and <111> direction. For <001> direction the flow stress is maximum for the 10% of *Cu* addition. When the strain value is around 0.27~0.30 for this direction, severe dislocation based activity takes place. The variation of dislocation density with strain further illustrates the stress-strain curve for <001> direction (see Fig. 2(a) and (d)). There are no dislocations up to a certain strain value where the first yielding takes place. After that for the pure *Al* in <001> dislocation density goes up. A similar trend is found for 5% *Al-Cu* alloy but the density of dislocation is lower than pure *Al*. The dislocation density is lowest for the 10% *Al-Cu* alloy though the stress-strain curve shows several peaks. From the dislocation density variation with strain, it is clear that the interactions between the dislocations annihilate themselves. This causes the peak formation in the stress-strain curve for the alloy. Overall, the addition of *Cu* causes lower dislocation density for <001> direction.

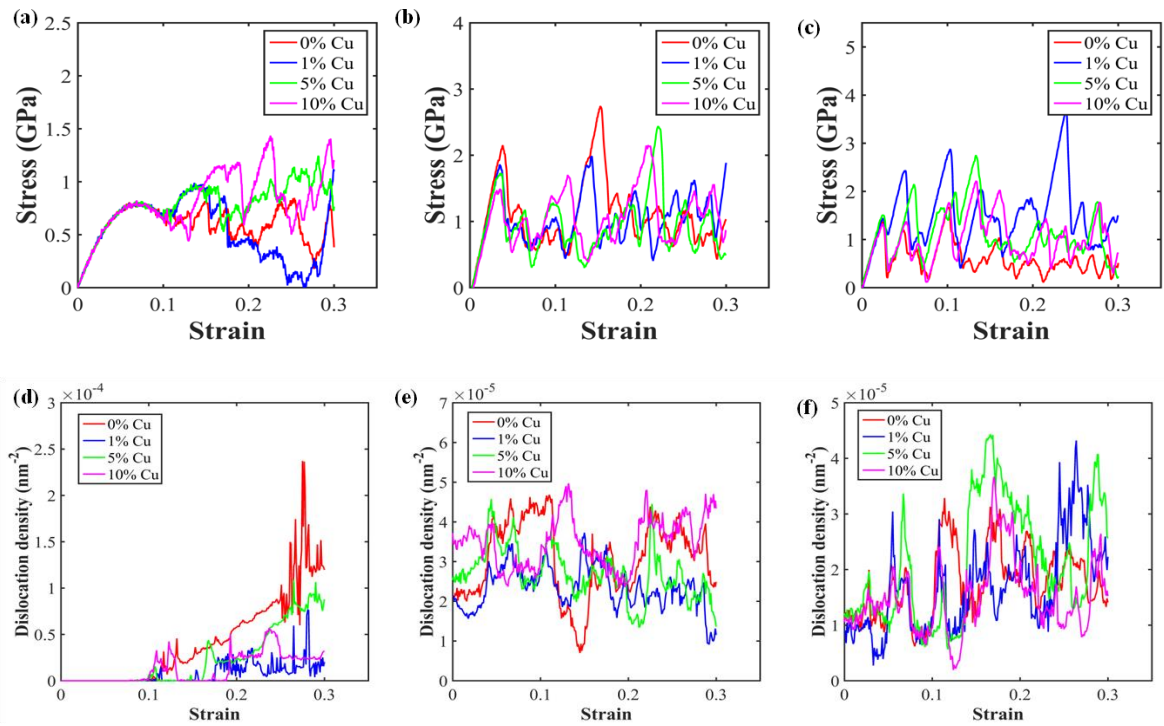

Fig 2: Compressive stress-strain Curve for *Al-Cu* alloy nanopillar for loading in (a) <001>, (b) <110> and (c) <111> direction for different *Cu* percentage. Variation in dislocation density for loading in (d) <001>, (e) <110> and (f) <111> direction for different *Cu* percentage.

The stress-strain responses for <110> direction are different from other directions of loading. The first yielding stress value goes down with the addition of *Cu* and 10% *Al-Cu* alloy shows the lowest value of first yielding stress. For <110> direction the highest peak is found for pure *Al* (0% *Al-Cu* alloy). The dislocation density plots with strain show that the density of dislocation drops significantly during these peak formation.

A similar trend is found for <111> direction loading but this time 1% *Al-Cu* alloy is showing the highest peaks and their dislocation density is reduced during these peak formation. However, the pure *Al* in <111> direction show the lowest yielding stress and flow stress. Therefore, <111> crystallographic direction is the weaker direction under compressive loading. The previous study of Wu et al. [25] also corroborates that <111> direction is weaker than <110> for loading. The dislocation nucleation and interactions are shown and discussed in the supplementary of this paper.

In Fig. 3 the alloying percentage effect on elastic and yielding properties are shown. The elastic modulus is higher for <111> direction for all alloying percentage and lower for <001> direction. The yielding stress is higher for the <110> direction and has a trend to decrease with higher *Cu* percentage. For <111> direction the yielding stress is higher for 5% *Al-Cu* alloy. The yielding stress is lower and does not vary with *Cu* addition for <001> direction. For the yielding strain, the first yielding occurs at a lower strain of around 0.025 for <111> direction. And the yielding strains for other two directions are comparable. Alloying element seems to have an insignificant effect on yielding strain.

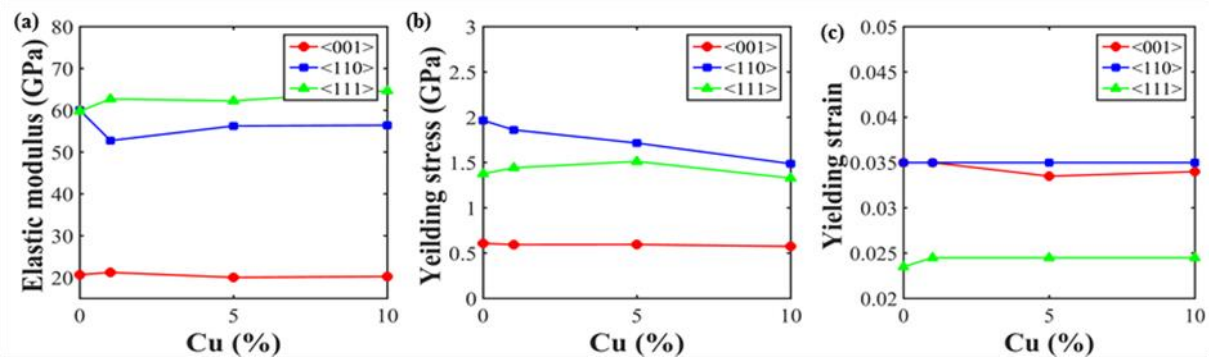

Fig.3: Variation of (a) elastic modulus, (b) yielding stress, (c) yielding strain with different *Cu* percentage of *Al-Cu* alloy.

## 4. Conclusions

The following conclusions can be listed from the results discussed:

- The peaks in the flow stress region of the stress-strain curves are the resultant of dislocation annihilation and this behavior has a strong relationship with loading direction and alloying percentage. For <001> direction higher alloying can sustain higher strength but <110> shows opposite trend. For <111> moderate alloying percentage have the highest strength.

- Elastic modulus and yielding stress are lower for <001> direction for all Cu percentage. Elastic modulus is highest for <111> direction but yielding stress is highest for <110> direction.
- Yielding strain has no significant effect on alloying percentage and highest for <110> and lowest for <111> direction.

**Acknowledgments:**

The author would like to express his gratitude to Department of Mechanical Engineering, BUET, Dhaka-1000 for providing the computational facilities. The author would like to acknowledge the contribution of Dr. Dibakar Datta, Dr. Mohammad Abdul Motalab and Dr. Monon Mahboob for their kind help in result analysis. The author is also thankful to the anonymous reviewer for his suggestions to improve this paper.

# Supplementary of "Molecular dynamics study of plasticity in *Al-Cu* alloy nanopillar due to compressive loading"

Satyajit Mojumder*

[a]Department of Mechanical Engineering, Bangladesh University of Engineering and Technology, Dhaka-1000, Bangladesh.

**Dislocation nucleation and propagation:**

The dislocation nucleation and propagation for different orientations and alloying percentages are shown in Figures S1-S12. In these figures, the dislocation formation during the first yielding, and after the first yielding are shown. The dislocation glide plane is different for the different orientation of loading. With the increment of alloying percentage, the primary slip plane does not change for <001> direction loading. But for <110> and <111> direction loading the slip plane changes with the different percentage of alloying elements. The dislocation interaction and propagation can be seen from these figures.

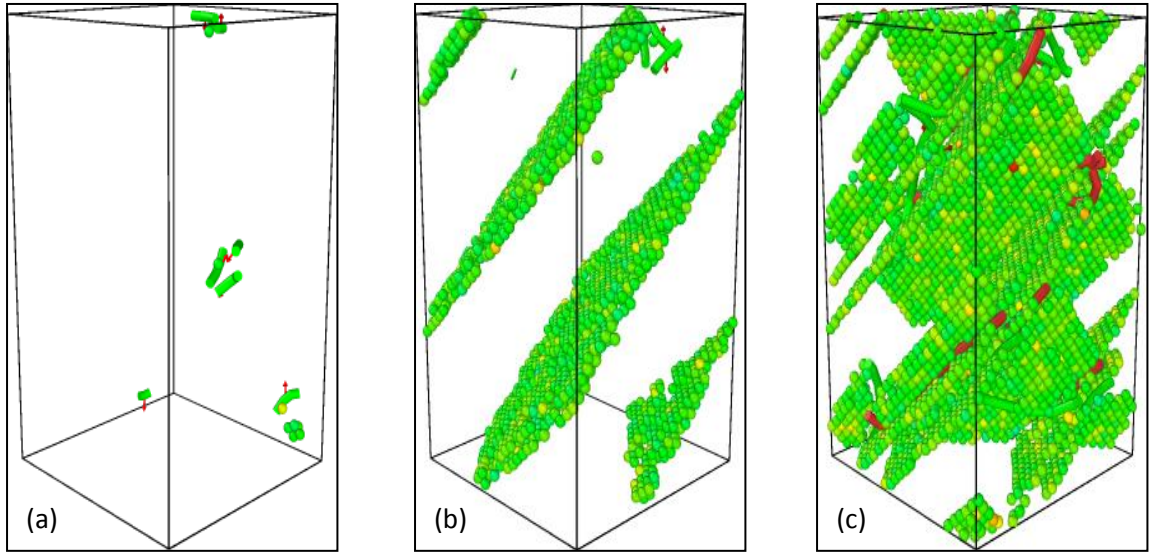

Fig. S1: Dislocation nucleation and propagation for <001> direction loading of Pure Aluminium crystal. (a) snapshot at strain =0.1, (b) snapshot at strain =0.11, (c) snapshot at strain =0.15. The snapshot are created using common neighbor analysis. The BCC and HCP structure are shown in the image which denotes the slip plane.

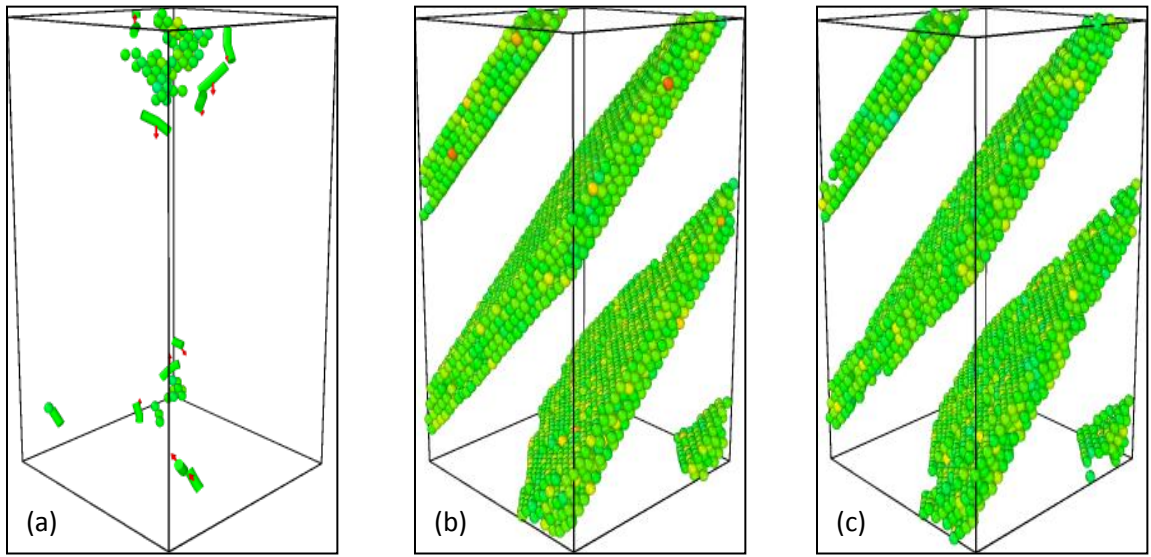

Fig. S2: Dislocation nucleation and propagation for <001> direction loading of Aluminium alloy with 1% Copper. (a) snapshot at strain =0.1, (b) snapshot at strain =0.12, (c) snapshot at strain =0.16. The snapshot are created using common neighbor analysis. The BCC and HCP structure are shown in the image which denotes the slip plane.

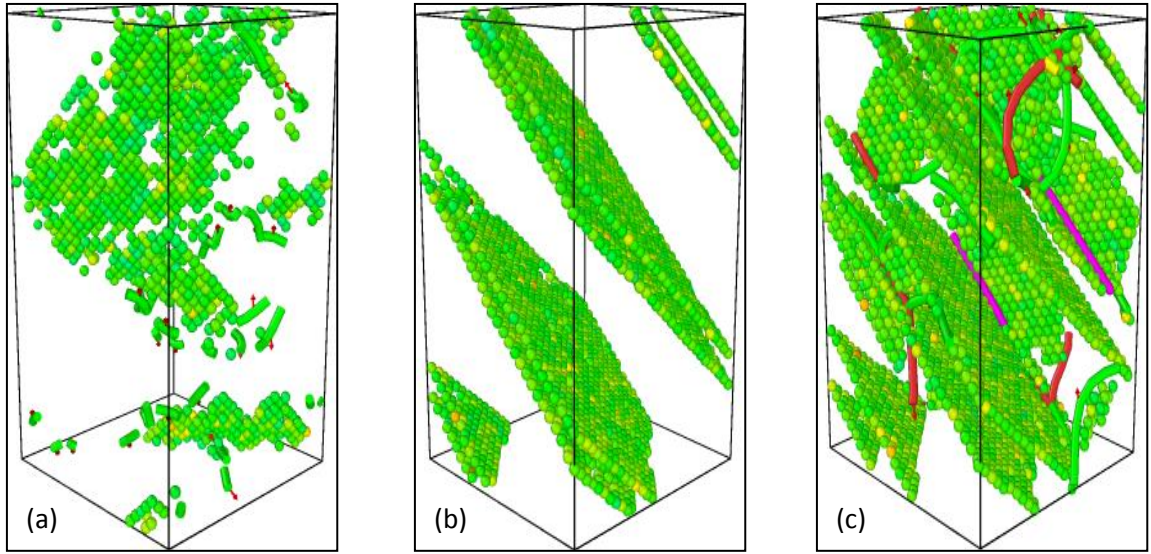

Fig. S3: Dislocation nucleation and propagation for <001> direction loading of Aluminium alloy with 5% Copper. (a) snapshot at strain =0.1, (b) snapshot at strain =0.12, (c) snapshot at strain =0.19. The snapshot are created using common neighbor analysis. The BCC and HCP structure are shown in the image which denotes the slip plane.

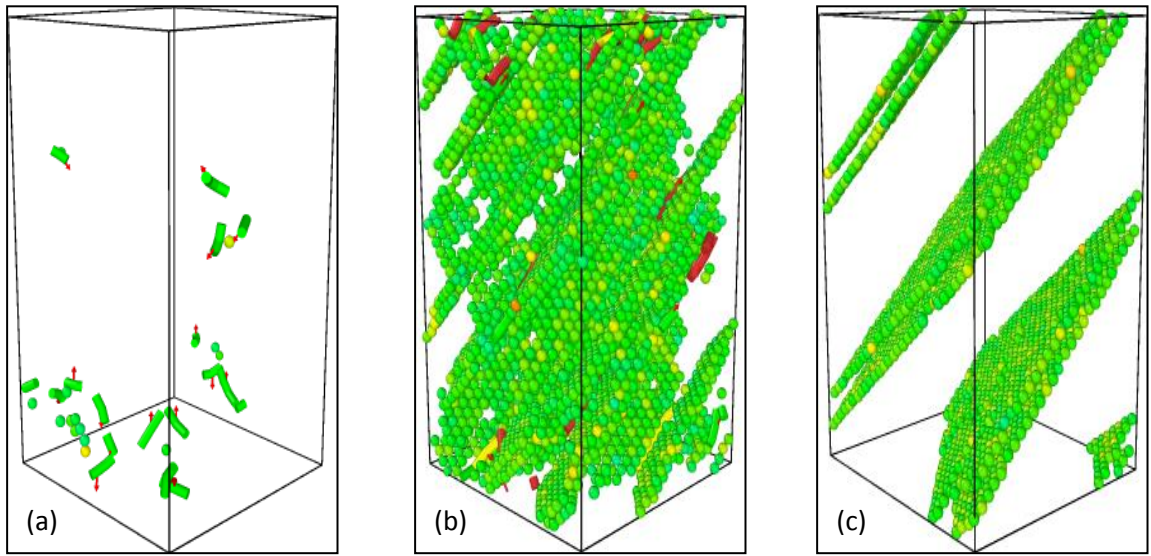

Fig. S4: Dislocation nucleation and propagation for <001> direction loading of Aluminium alloy with 10% Copper. (a) snapshot at strain =0.1, (b) snapshot at strain =0.12, (c) snapshot at strain =0.16. The snapshot are created using common neighbor analysis. The BCC and HCP structure are shown in the image which denotes the slip plane.

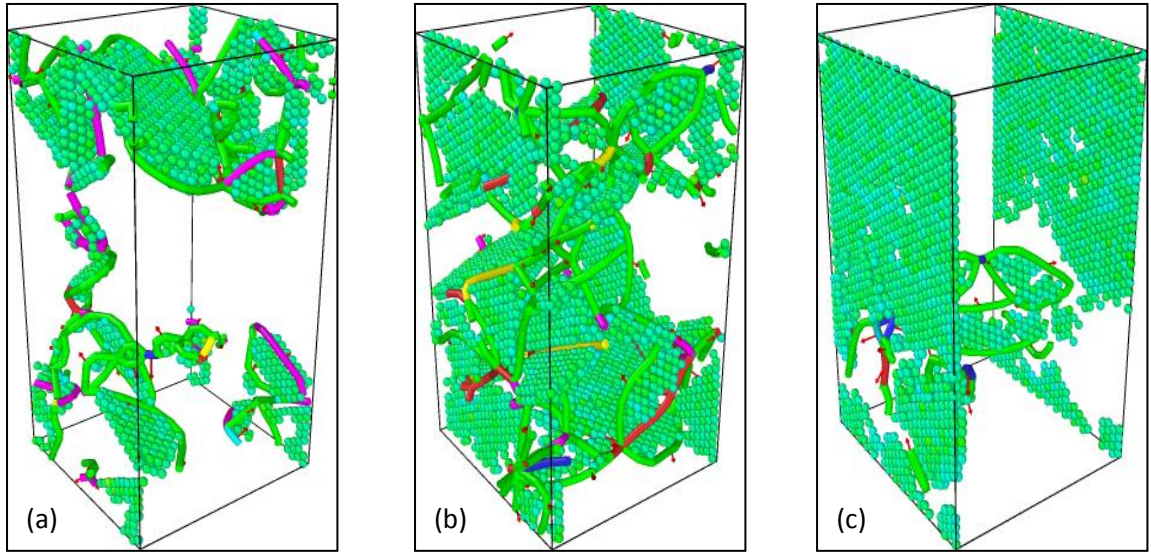

Fig. S5: Dislocation nucleation and propagation for <001> direction loading of Pure Aluminium crystal. (a) snapshot at strain =0.04, (b) snapshot at strain =0.1, (c) snapshot at strain =0.15. The snapshot are created using common neighbor analysis. The BCC and HCP structure are shown in the image which denotes the slip plane.

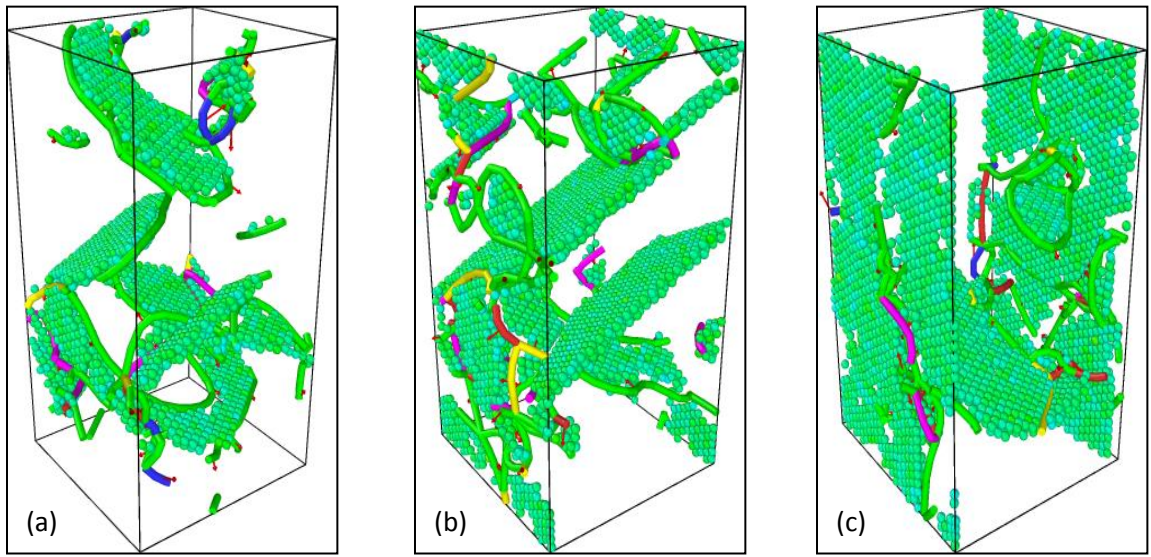

Fig. S6: Dislocation nucleation and propagation for <110> direction loading of Aluminium alloy with 1% Copper. (a) snapshot at strain =0.04, (b) snapshot at strain =0.077, (c) snapshot at strain =0.14. The snapshot are created using common neighbor analysis. The BCC and HCP structure are shown in the image which denotes the slip plane.

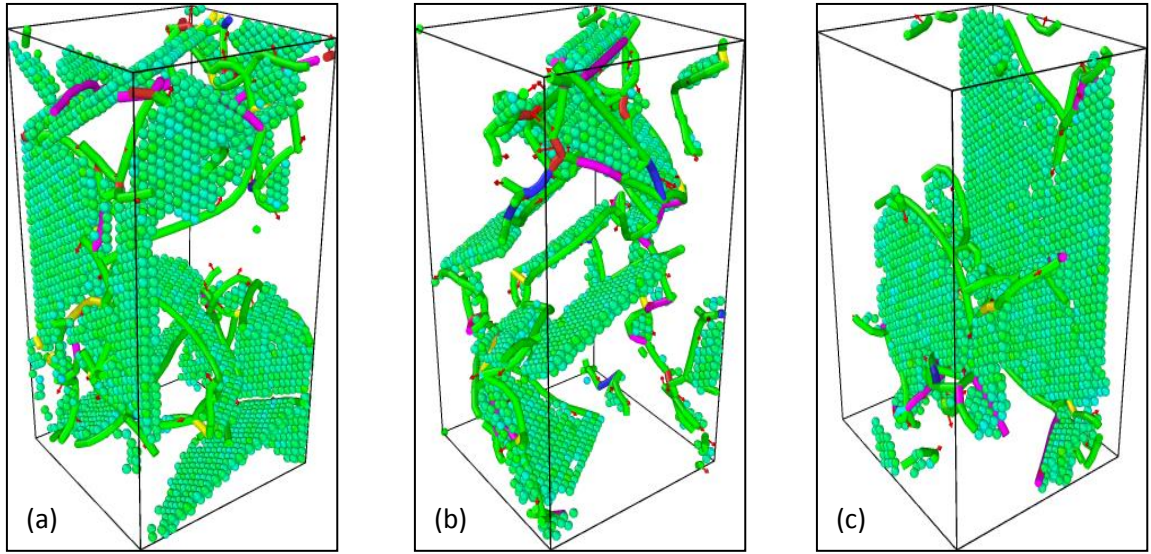

Fig. S7: Dislocation nucleation and propagation for <110> direction loading of Aluminium alloy with 5% Copper. (a) snapshot at strain =0.04, (b) snapshot at strain =0.07, (c) snapshot at strain =0.14. The snapshot are created using common neighbor analysis. The BCC and HCP structure are shown in the image which denotes the slip plane.

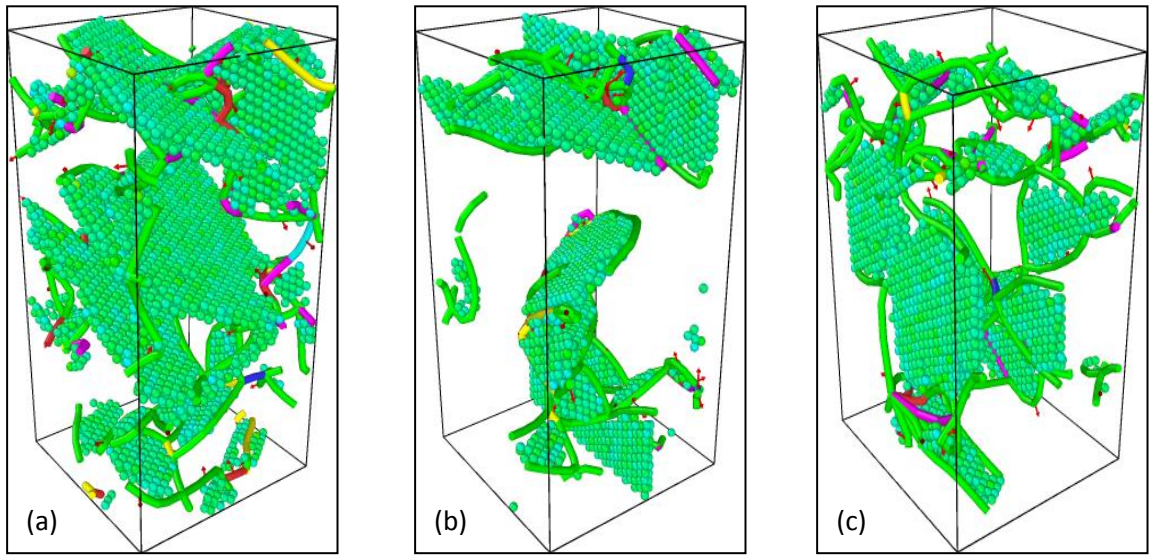

Fig. S8: Dislocation nucleation and propagation for <110> direction loading of Aluminium alloy with 10% Copper. (a) snapshot at strain =0.04, (b) snapshot at strain =0.065, (c) snapshot at strain =0.148. The snapshot are created using common neighbor analysis. The BCC and HCP structure are shown in the image which denotes the slip plane.

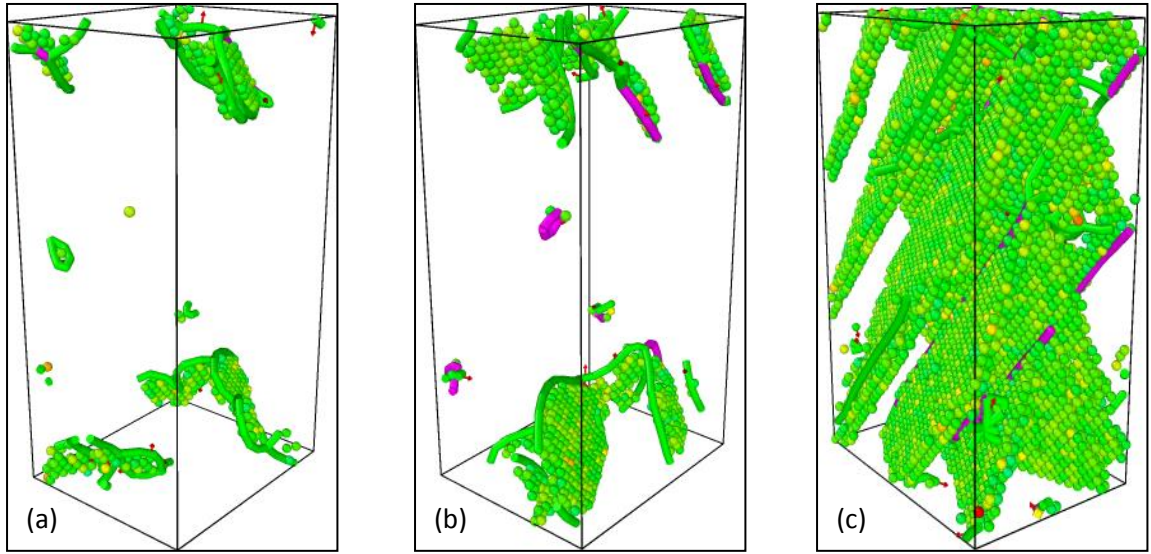

Fig. S9: Dislocation nucleation and propagation for <111> direction loading of Pure Aluminium crystal. (a) snapshot at strain =0.025, (b) snapshot at strain =0.06, (c) snapshot at strain =0.11. The snapshot are created using common neighbor analysis. The BCC and HCP structure are shown in the image which denotes the slip plane.

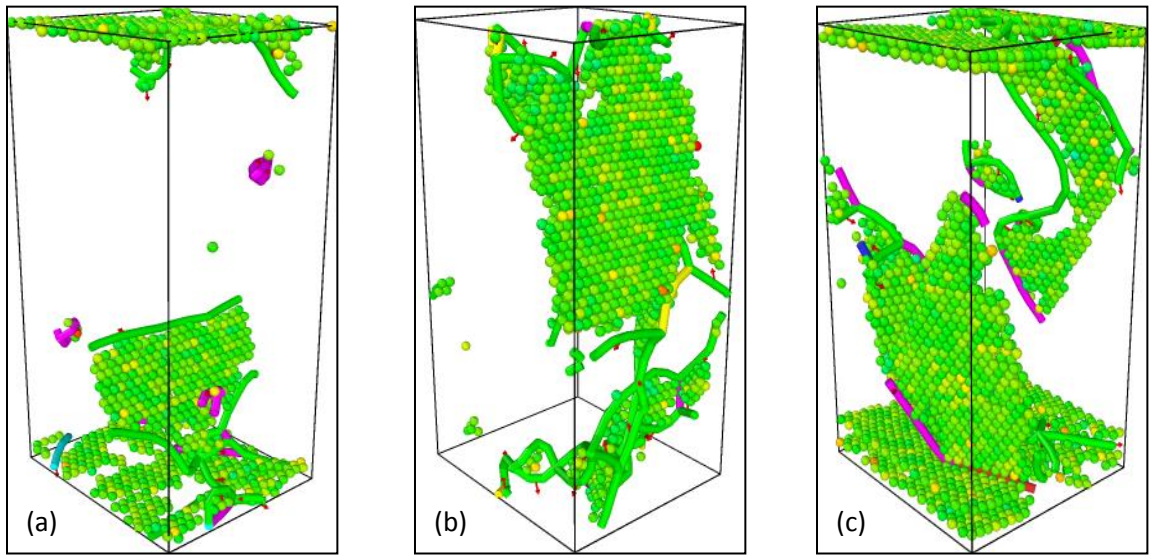

Fig. S10: Dislocation nucleation and propagation for <111> direction loading of Aluminium alloy with 1% Copper. (a) snapshot at strain =0.025, (b) snapshot at strain =0.058, (c) snapshot at strain =0.113. The snapshot are created using common neighbor analysis. The BCC and HCP structure are shown in the image which denotes the slip plane.

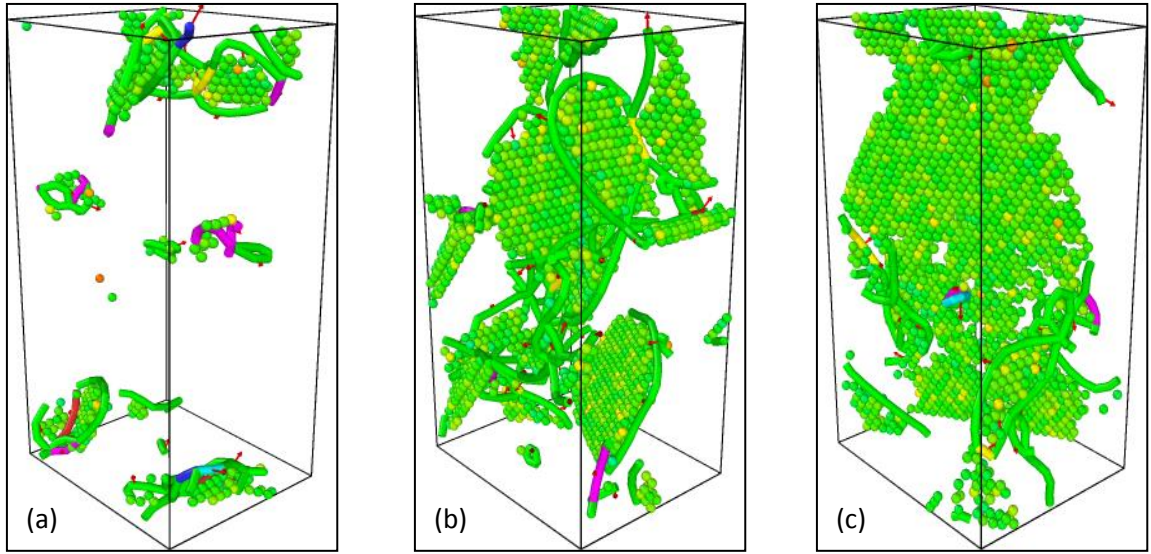

Fig. S11: Dislocation nucleation and propagation for <111> direction loading of Aluminium alloy with 5% Copper. (a) snapshot at strain =0.025, (b) snapshot at strain =0.067, (c) snapshot at strain =0.112. The snapshot are created using common neighbor analysis. The BCC and HCP structure are shown in the image which denotes the slip plane.

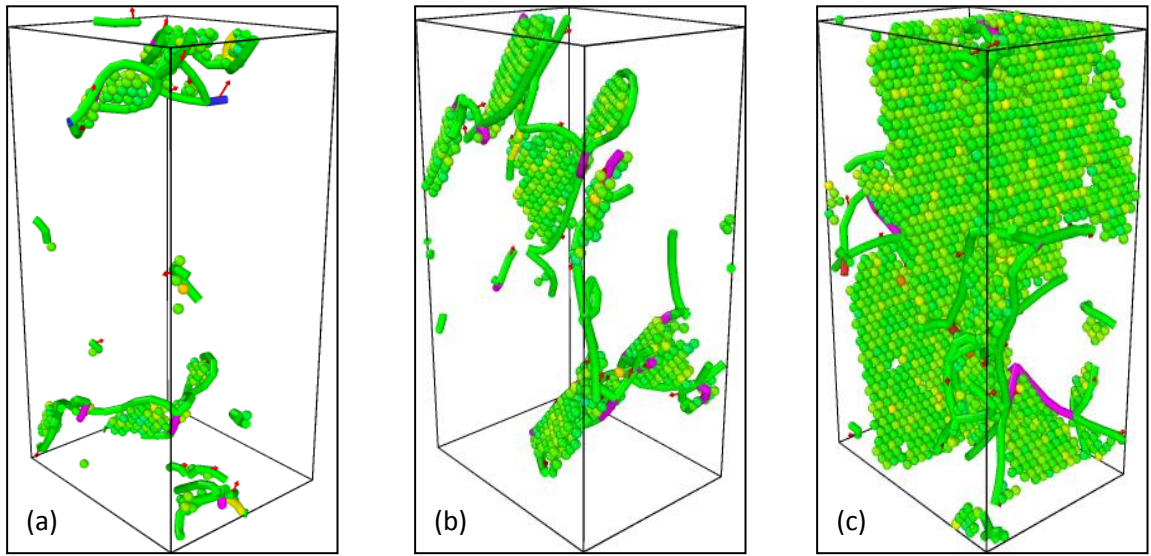

Fig. S12: Dislocation nucleation and propagation for <111> direction loading of Aluminium alloy with 10% Copper. (a) snapshot at strain =0.025, (b) snapshot at strain =0.072, (c) snapshot at strain =0.112. The snapshot are created using common neighbor analysis. The BCC and HCP structure are shown in the image which denotes the slip plane.